\documentclass[sigconf]{acmart}
\copyrightyear{2024}
\acmYear{2024}
\setcopyright{acmlicensed}\acmConference[FDG 2024]{Proceedings of the 19th International Conference on the Foundations of Digital Games}{May 21--24, 2024}{Worcester, MA, USA}
\acmBooktitle{Proceedings of the 19th International Conference on the Foundations of Digital Games (FDG 2024), May 21--24, 2024, Worcester, MA, USA}
\acmDOI{10.1145/3649921.3650017}
\acmISBN{979-8-4007-0955-5/24/05}
\AtBeginDocument{%
  \providecommand\BibTeX{{%
    \normalfont B\kern-0.5em{\scshape i\kern-0.25em b}\kern-0.8em\TeX}}}
    
\usepackage{tikz}
\usepackage{subcaption}
\usepackage{natbib}
\usepackage{listings}
\usepackage{mathtools}
\usepackage{quoting}
\quotingsetup{vskip=5pt}
\usepackage{balance}

\usepackage{amsthm}

\setcopyright{acmcopyright}
\copyrightyear{2022}
\acmYear{2022}
\acmDOI{XXXXXXX.XXXXXXX}

\usepackage[clock,alpine]{ifsym}

\acmConference[FDG '24]{Make sure to enter the correct
  conference title from your rights confirmation email}{April 11--14,
  2023}{Lisbon, Portugal}
\copyrightyear{2024}
\acmYear{2024}
\setcopyright{acmlicensed}\acmConference[FDG 2024]{Proceedings of the 19th International Conference on the Foundations of Digital Games}{May 21--24, 2024}{Worcester, MA, USA}
\acmBooktitle{Proceedings of the 19th International Conference on the Foundations of Digital Games (FDG 2024), May 21--24, 2024, Worcester, MA, USA}
\acmDOI{10.1145/3649921.3650017}
\acmISBN{979-8-4007-0955-5/24/05}

\usepackage{xcolor}

\newcommand*{\mathabxbfamily}{\fontencoding{U}\fontfamily{mathb}\selectfont}
\DeclareFontFamily{U}{mathb}{\hyphenchar\font45}
\DeclareFontShape{U}{mathb}{m}{n}{
      <5> <6> <7> <8> <9> <10> gen * mathb
      <10.95> mathb10 <12> <14.4> <17.28> <20.74> <24.88> mathb12
      }{}

\newcommand*{\Neptune}{{\text{\mathabxbfamily\char"48}}}

\makeatletter
\def\@fnsymbol#1{\ensuremath{
\ifcase#1\or
\Small{\text{\Mountain}} \or  
\Neptune \or 
\mathsection \or
\mathparagraph \or
\|\or **
\or
\dagger\dagger
\or \ddagger\ddagger
\else\@ctrerr\fi}}
\makeatother

\renewenvironment{quote}{%
  \list{}{%
    \leftmargin0.5cm   
    \rightmargin\leftmargin
  }
  \item\relax
}
{\endlist}

\begin{document}

\title{How To Save A World: The Go-Along Interview as Game Preservation Methodology in Wurm Online}



\author{Florence Smith Nicholls}
\affiliation{%
  \institution{Queen Mary University of London}
  \city{London}
  \country{UK}
}
\email{florence@knivesandpaintbrushes.org }

\authornote{This author should be cited with their full surname "Smith Nicholls" and they/them pronouns.}

\author{Michael Cook}
\affiliation{%
  \institution{King's College London}
  \city{London}
  \country{UK}}
\email{mike@possibilityspace.org}

\authornote{References to this author may be made using the he/him masculine or they/them singular neutral pronouns.}

\renewcommand{\shortauthors}{Smith Nicholls and Cook}

\begin{abstract}
Massively multiplayer online (MMO) games boomed in the late 1990s to 2000s. In parallel, ethnographic studies of these communities emerged, generally involving participant observation and interviews. Several decades on, many MMOs have been reconfigured, remastered or are potentially no longer accessible at all, which presents challenges for their continued study and long-term preservation. In this paper we explore the “go-along” methodology, in which a researcher joins a participant on a walk through a familiar place and asks them questions, as a qualitative research method applicable for the study and preservation of games culture. Though the methodology has been introduced in digital media studies, to date it has had limited application in digital games, if at all. We report on a pilot study exploring applications of the go-along method to the sandbox MMO \textit{Wurm Online}; a persistent, player-directed world with a rich history. We report on our motivations for the work, our analysis of the resulting interviews, and our reflections on both the use of go-alongs in digital games, as well as the unique and inspiring culture and community of this lesser-known game.

\end{abstract}




\maketitle

\section{Introduction}
The ethnographic study of massive multiplayer online games (hereafter MMOs) developed into a thriving field in the mid-2000s. A key characteristic that has made MMOs so attractive to ethnographers is their apparent persistence \cite{boellstorff} -- the worlds continued to exist and change regardless if particular players were online. The deep irony of this is that the continued persistence of MMOs is very fragile, relying not only on server architecture but sustained communities to cultivate their culture.
From a games preservation point of view, maintaining access to the MMO software is profoundly insufficient. Records need to be made of what it is actually like to experience an MMO, with all its attendant social and cultural complexities. Ethnography is one method which has been employed to do this, however to date the “go-along” method has had limited application in games, if at all. With this methodology, researchers accompany participants on a walk, asking them questions and reflecting on the environment that they pass through. We wanted to explore the potential of this method for capturing a deeper understanding of player experience and ties to an MMO environment.
We chose to conduct go-alongs in \textit{Wurm Online}\cite{wurm}, an MMO with a small but consistent player base which has yet to be subject to considerable academic study. Wurm presented an interesting case study for go-alongs as it has been online since 2006, with player-created structures over a decade old that still exist within it. 

\section{Background}
\subsection{Ethnographic study of MMOs}

The ethnographic study of MMOs builds on earlier work into the anthropology of text-based online multiplayer games, such as MUDs (multi-user domains), which provided an opportunity to understand how culture and communities formed within these emerging digital contexts \cite{mayra2013}, and the importance of not just spectating but participating in play as a researcher \cite{mortensen}. Two key strands in the work on ethnographic studies of MMOs relate to methodology; what does it mean to apply the ethnographic method in a digital space, and to what extent is the study of MMOs limited by  constructing binary dichotomies between the physical and digital?

In one of the classic digital ethnographies of the 00’s, \textit{Coming of Age in Second Life}, Boellstorff contends that he can study virtual worlds “on their own terms,” conducting his research entirely within Second Life itself \cite{boellstorff2008}. This fits into the “third ethnographic scale” in Boellstroff’s own typology, in which the first ethnographic scale interfaces between the real and the virtual, and the second across two or more virtual worlds \cite{boell2010}. Though T.L. Taylor’s ethnographic work on the MMO \textit{EverQuest} and an associated fan event \cite{taylor2006} arguably fits the first scale, and Pearce’s ethnography of a displaced MMO community \cite{pearce2009} fits the second, Nardi’s work on \textit{World of Warcraft} \cite{nardi2010} involves in-person interviews of players accessing the game, which potentially problematises these categories. 

A key criticism of Boellstorff’s work is that he limits his analysis exclusively to the virtual world, not considering the lived reality of players \cite{heuser2019}. This criticism ties in with a long-running theme in ethnographic studies of MMOs: the fallacy of the virtual/real dichotomy. Nick Taylor \cite{nicktaylor2008} has claimed that ethnographic research into MMOs has followed a trend of reifying the online/offline demarcation, erasing the presence of the ethnographer in their own analysis, a process which he terms “periscopic play.” Taylor also argues that if classic ethnographic methods are transposed onto MMOs uncritically, then this new wave of scholarship runs the risk of emulating the colonial narrative of the ‘lone ethnographer’ who treats digital space as just another ‘pristine wilderness’ without accountability to their research subjects. More recent work continues the trend of complicating the role of the ethnographer, with Wilde using autoethnography to reflect on the post-human entanglements she had with her \textit{World of Warcraft} avatar \cite{wilde2018}.

Interestingly, in a 2022 paper T.L. Taylor draws on a piece of classic ethnography, Geertz’s account of deep play and the Balinese cock fight \cite{geertz}, to make the point that ethnography is inherently playful and that we should look to “qualities of ethnographic practice that are not easily distilled to work that looks serious, planned or controlled” \cite{taylor2022}. The go-along methodology that we use in this paper arguably fits into Boellstorff’s third ethnographic scale, however as a hybrid of participant observation and interview, it does not seek to replicate the classic ethnographic model. The affordances of the go-along, as a method that invites participants to lead the conversation according to their everyday experiences in a particular place, led us to look beyond the “serious, planned or controlled,” and reflect on our role as researchers invited to share in that space.

\subsection{Archiving MMOs}
While ethnographic methods can aid in the understanding of MMO play cultures, they can also aid in preserving them as well. Game preservation in general is fraught with numerous complications, not least access to original software and hardware, but MMOs have the added complication of being online play experiences which are contingent on the existence of a sustained community. MMOs like \textit{Wurm Online} depend on the maintenance of server infrastructures, while ongoing software updates also have the potential to change the game landscape irrevocably \cite{barbier2022}.

Bartle \cite{bartle_2015} observes that the application of both archaeological and anthropological methodologies is required for the preservation of MMO worlds. In the case of the former, there has been work on archaeologically recording the remains of settlements abandoned after a software patch in \textit{No Man’s Sky} \cite{reinhard2021}. Johnson's ethnographic work on player engagement with the past in \textit{Skyrim} combined both participant observation and an online survey  \cite{johnson}. Hansen has used both archaeological site mapping and ethnographic interviews to investigate a player community in \textit{Star Wars Galaxies} \cite{hansen2022}. Pearce’s work \cite{pearce2009} on the \textit{Uru: Ages of Myst} community following the closure of its servers provides documentation of how fragile and ephemeral these MMO cultures can be. 

A key concern in the academic literature on preserving MMOs is the need to record the experience of play and not just maintain access to the game worlds themselves. There are numerous strategies for this, such as archiving live streams or other video footage \cite{morris2023}, though:

\begin{quoting} 
\noindent Ethnographic narratives can present the subjective and emotional experience of playing an MMO in a way that gameplay videos cannot. Anyone aiming to document the player community of an MMO should consider the ethnographic narrative an essential tool.\cite{murphy2015}
\end{quoting}

\noindent Leaning into the subjectivity of the ethnographic account can be one way of avoiding the potential decontextualization of the lived experience of MMO players through their “heritigazation” in the archive \cite{buccitelli2022}. The go-along, informed not only by the contemporary experience of the MMO landscape, but its attendant memories, is one way of constructing this kind of emotional ethnographic narrative.

\subsection{Wurm Online}
\textit{Wurm Online} is an MMO originally launched in 2006. Unlike more story-focused MMOs, \textit{Wurm} is an open-ended sandbox game, where players are largely given no direction, and instead set their own goals to achieve. A player has hundreds of skills which govern both the chance of succeeding at certain actions and the quality and performance with which those actions are carried out. Some skills are more general, such as `Mining' while other skills are more specific -- for example, the `Cooking' category has five sub-skills including Baking and Butchering. Skill gain is notoriously slow -- only a handful of players in the game's history have reached the maximum level of 100 for even a single skill. 

A key feature of\textit{ Wurm Online's} design is that the landscape can be modified by any player, and that these changes persist. Mining rock, digging holes and chopping down trees all have a permanent impact on the geography of the game world. Player-made structures and items also persist in the world, but unlike terraforming actions these changes slowly decay over time: buildings degrade and parts will eventually collapse, while items are subject to wear and tear and will disappear. These processes are slow; some structures can last for several (real-world) years without maintenance. Every item also bears on it the signature of who made it, although it may not be readable if the item is of low quality.

\textit{Wurm Online} is run as a free-to-play game, however players can purchase in-game currency, one function of which is to pay the upkeep on a `deed', which designates them as the owner of a small area of land. Structures built within a deed do not decay as long as a regular payment is made, and deed owners can set permissions to limit what actions other players can take on the deed. \textit{Wurm Online} currently operates seventeen servers, each of which is modelled as a unique island with its own landscape, with some servers being clustered together. The oldest active server, Independence, has been in operation since 2009, while the newest, Cadence, opened in 2020. Some servers have particular modifications attached to them (such as allowing \textit{player versus player} combat). Players can travel between servers within a cluster, however until very recently this was a difficult and time-consuming task and beyond the reach of newer players.

\subsection{The Dragon Fang Pass}
Players often collaborate on large projects that require skilled characters and large amounts of time and resources. These projects vary in purpose, but include the creation of infrastructure such as highways and canals, and the construction of monuments and feats of engineering. During an initial exploration of \textit{Wurm's} Independence server we encountered the \textit{Dragon Fang Pass}, a large tunnel dug through the largest mountain in Wurm, the eponymous Dragon Fang. This tunnel became the starting point for our investigation and conversations with the community. 


\subsection{Archiving Wurm}
\textit{Wurm's} community is highly active, and many members have been playing for over ten years and have a deep connection to the community, the world and the game. Many player activities constitute a kind of archival work, either consciously or otherwise, that contribute to a partial preservation of the culture and history of its players. Some of this work emerges from necessity, such as its player-made maps, which can be traced back over a decade \cite{wurmmaps}. Since \textit{Wurm} does not have a meaningful in-game map, and the world is full of buildings, roads and cultural sites players have added, these player-made maps are both functional for active players and act as an archive of how the world and its players have changed over time. Other projects are social, rather than functional -- many videos exist documenting player activity, from time-lapses of monuments being constructed, to everyday video diaries of player life in the game.

The changing nature of online communications has affected the \textit{passive} archiving of \textit{Wurm's} community. Early on in the game's development \textit{Wurm} players communicated often through IRC, an instant messaging protocol equivalent to a live chat. Over time, the \textit{Wurm} community transitioned to the use of online message boards. This format creates a static and public record of discussions, meaning that one can still view conversations about \textit{Wurm} and its culture today by looking at the forum archives. In recent years, social networking app Discord has grown as a second place for \textit{Wurm's} community. However, from a preservation perspective Discord has two key disadvantages compared to forums: discussions cannot be easily viewed or archived publicly; and the server itself is not owned by the community. Discord's inaccessibility and lack of archival tools poses serious problems for games preservationists, for \textit{Wurm} and beyond. 

\textit{Wurm} players also engage in active, \textit{intentional} archival work through the creation of shared resources. The \textit{Wurmpedia}, a wiki maintained by the community, is created in the style of many game wikis, to provide a catalogue of knowledge about the game to assist players \cite{wurmpedia}. However, unlike a wiki for a more static game, the \textit{Wurmpedia} also records historical and cultural knowledge about the player base. Similar projects seek to archive information from the game for other reasons. For example, a player called Andrea maintains a fan site of craftable item 3D models. The aim of the project is to make it "easier to decide which items to create for your homes and deeds to get the right look" \cite{andrea}. 

Some archival work takes place in the game itself. The Rockcliff Museum, which opened in 2022, is an in-game museum curated by a player known as Nirav in order to recognise, record and preserve many different aspects of community culture \cite{rockcliff}. \textit{Wurm} also has a formal notion of archival and preservation, in the form of Heritage Sites. Players can submit applications to the developers to request a location be granted heritage status, which confers special protections that prevent players from modifying or damaging the structure in question. 




\section{Methodology}
\subsection{Research questions}
In this study, we were interested in the affordances of the go-along in a digital context, specifically with regards to how it could elicit responses based on presence in particular in-game locations, and how players conceived of \textit{Wurm’s} heritage.  As such, we formed the following research questions:
\begin{itemize}
\item RQ1: How did the go-along methodology contribute to our understanding of the player experience of \textit{Wurm}?

\item RQ2: How are player memories tied to specific locations?

\item RQ3: How do players approach the curation of their own history and that of the game more broadly?
\end{itemize}

\begin{figure*}[h!]
\centering
\hfill
\begin{subfigure}[b]{0.45\textwidth}
    \includegraphics[width=\textwidth]{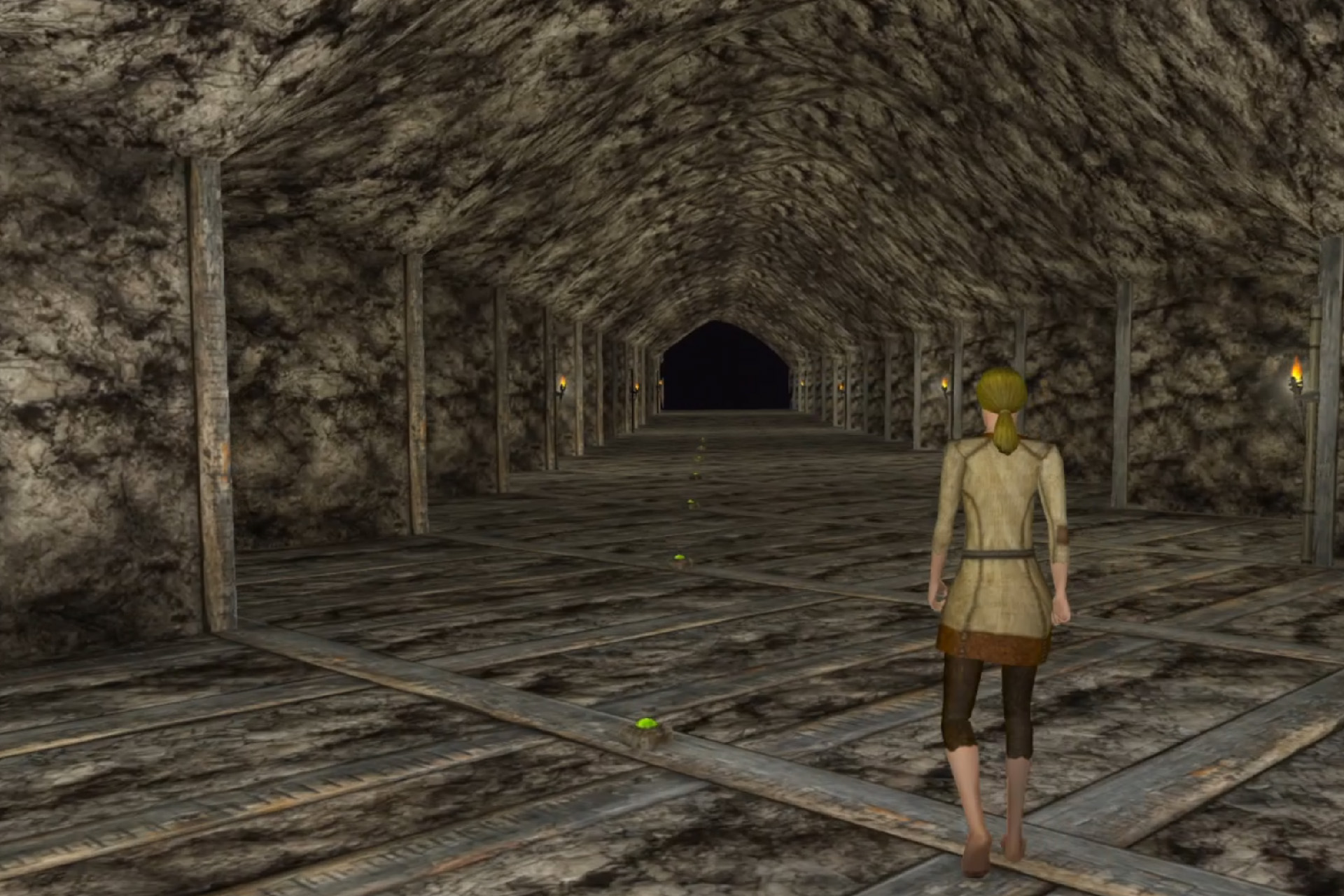}
    \caption{R1 (pictured) and R2 walk through the Dragon Fang Pass on their way to meet Participant 1 (Nirav) at the Rockcliff Museum for the go-along. In total, the authors walked the pass nine times in the course of this work.}
    \label{fig:wurm_screenshot_5}
\end{subfigure}
\hfill
\begin{subfigure}[b]{0.45\textwidth}
    \includegraphics[width=\textwidth]{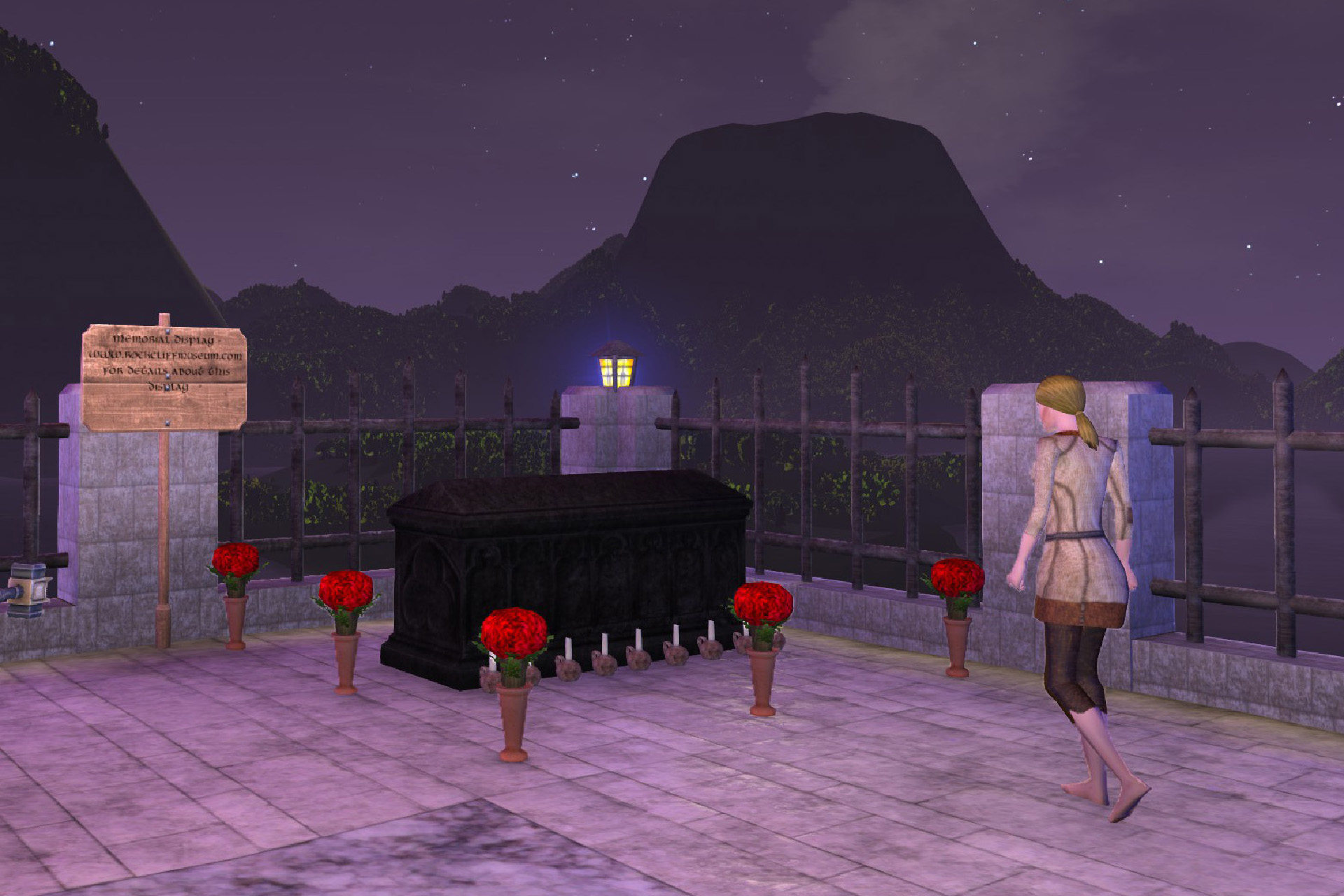}
    \caption{A memorial to players who have passed away in real life, as part of the Rockcliff Museum. Participant 1: ``I felt like [this] corner was a great place `cos there's kinda this feeling of things going on infinitely... in this view here'.}
    \label{fig:wurm_screenshot_2}
\end{subfigure}
\hfill
\\
\hfill
\begin{subfigure}[b]{0.45\textwidth}
    \includegraphics[width=\textwidth]{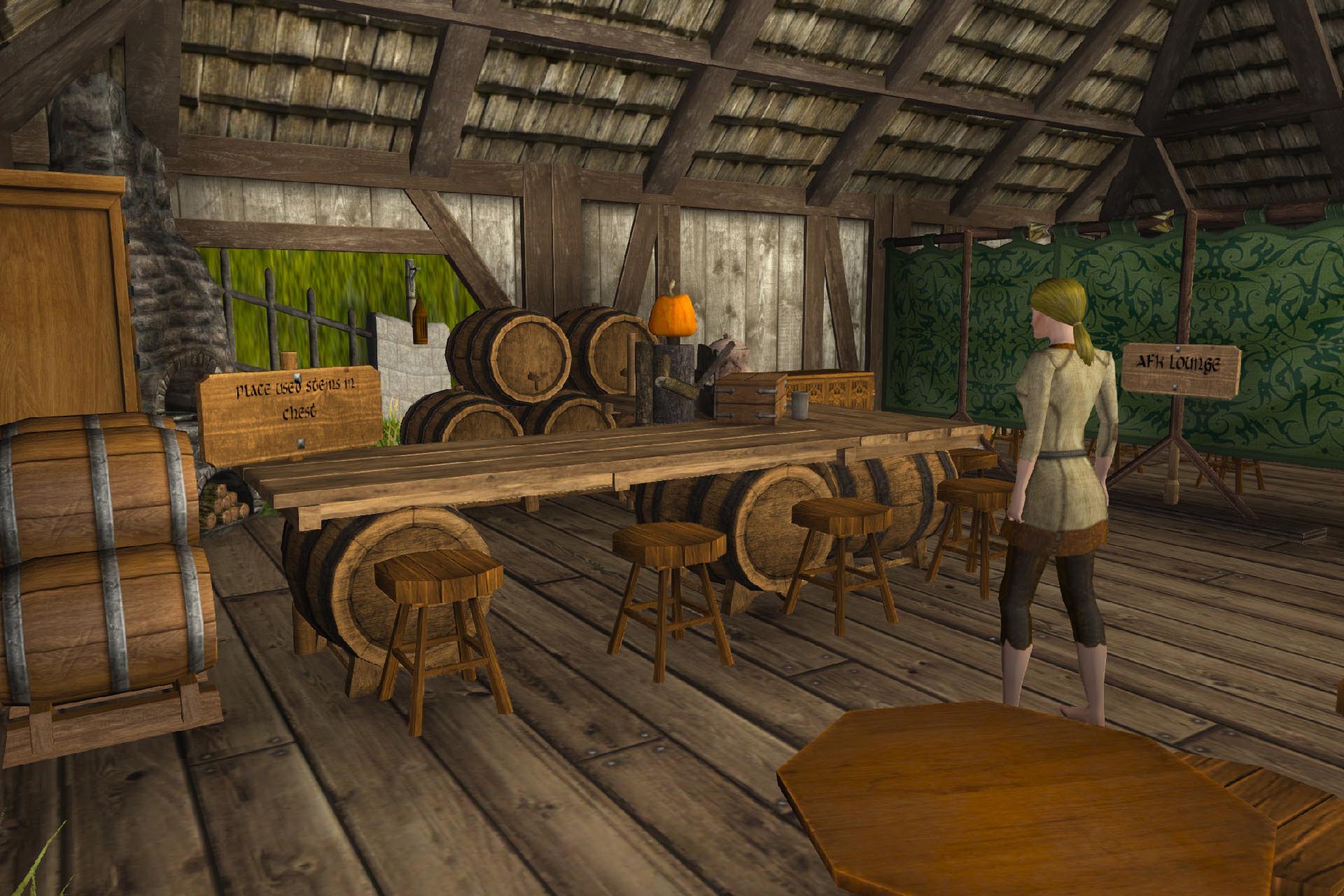}
    \caption{R1 stands in the tavern exhibit in the museum. This was a temporary construction built by a player for the museum's opening gala. This player is well-known for building and operating taverns at important in-game events, and so once the gala was finished it became a permanent record of their work.}
    \label{fig:wurm_screenshot_3}
\end{subfigure}
\hfill
\begin{subfigure}[b]{0.45\textwidth}
    \includegraphics[width=\textwidth]{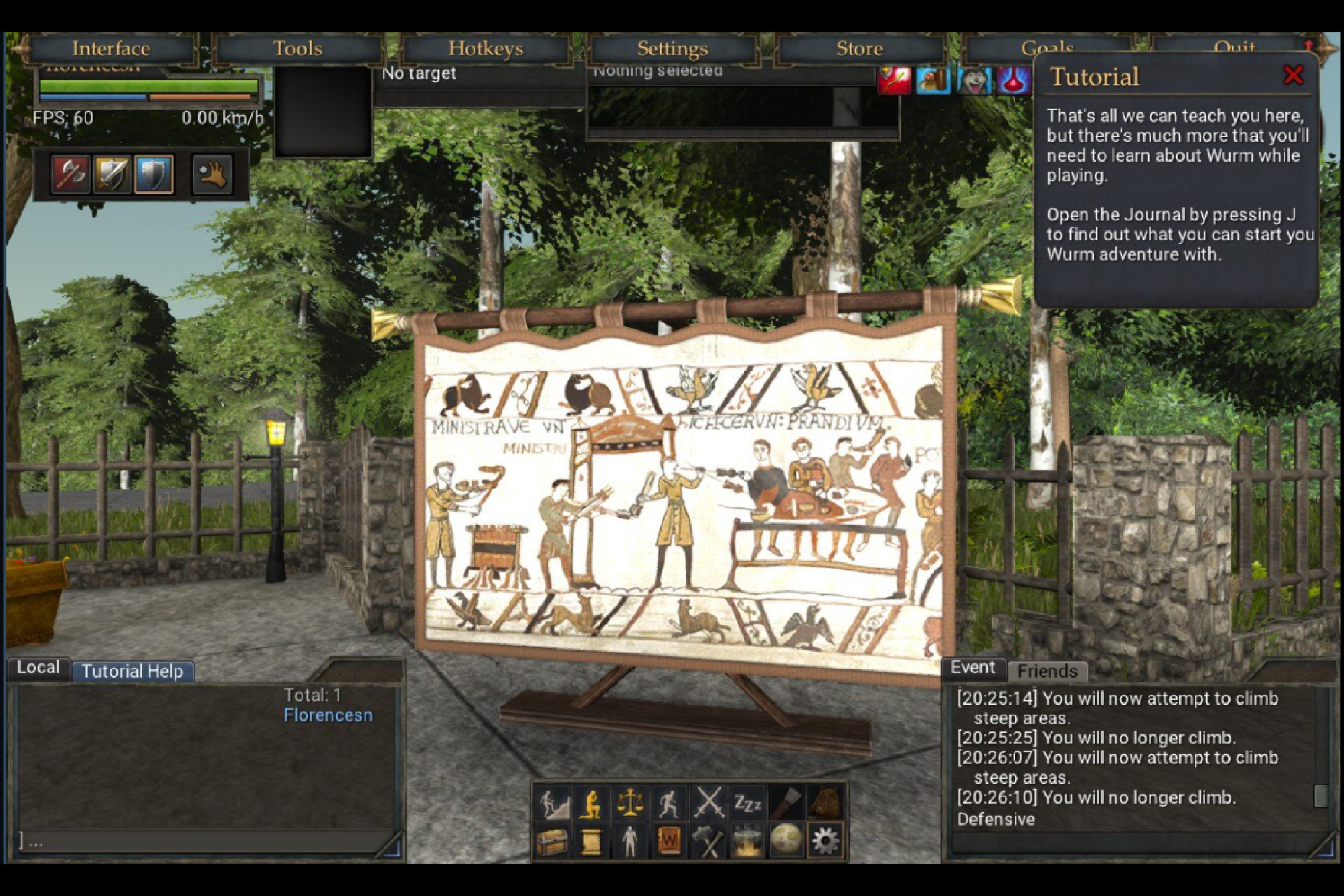}
    \caption{A screenshot taken during R1's playthrough of the game's tutorial. In this screenshot the full default user interface for the game is visible (as viewed on a Steam Deck). The action log is in the bottom right, chat is in the bottom left, while a hotbar for actions is in the bottom center.}
    \label{fig:wurm_screenshot_4}
\end{subfigure}
\hfill
\\
\hfill
\begin{subfigure}[b]{0.45\textwidth}
    \includegraphics[width=\textwidth]{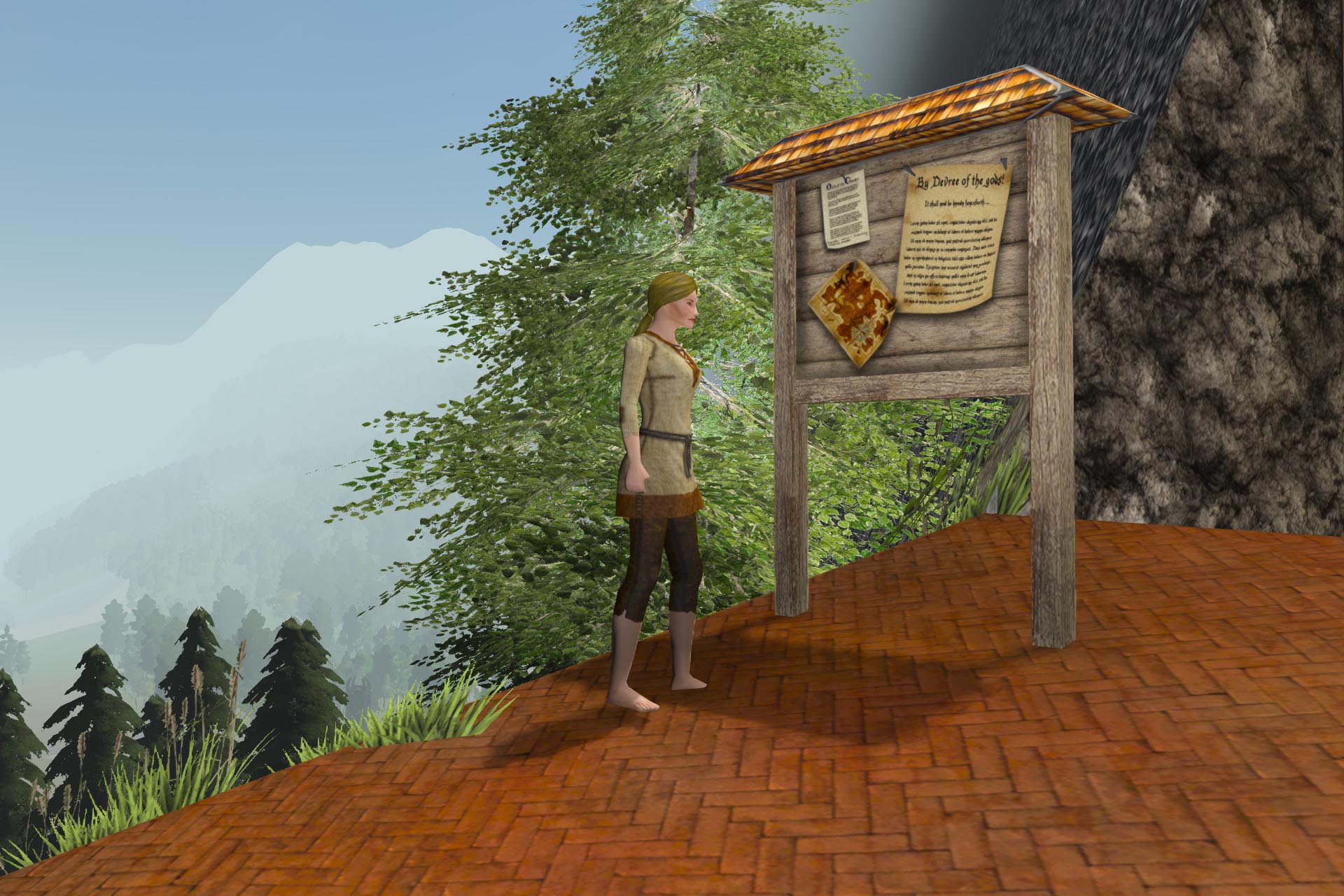}
    \caption{R1 standing in front of a sign marking the Dragon Fang Pass. It reads: "Dragon Fang Pass - Heritage Site - Do not alter without consent of Wurm GM team."} 
    \label{fig:wurm_screenshot_1}
\end{subfigure}
\hfill
\begin{subfigure}[b]{0.45\textwidth}
    \includegraphics[width=\textwidth]{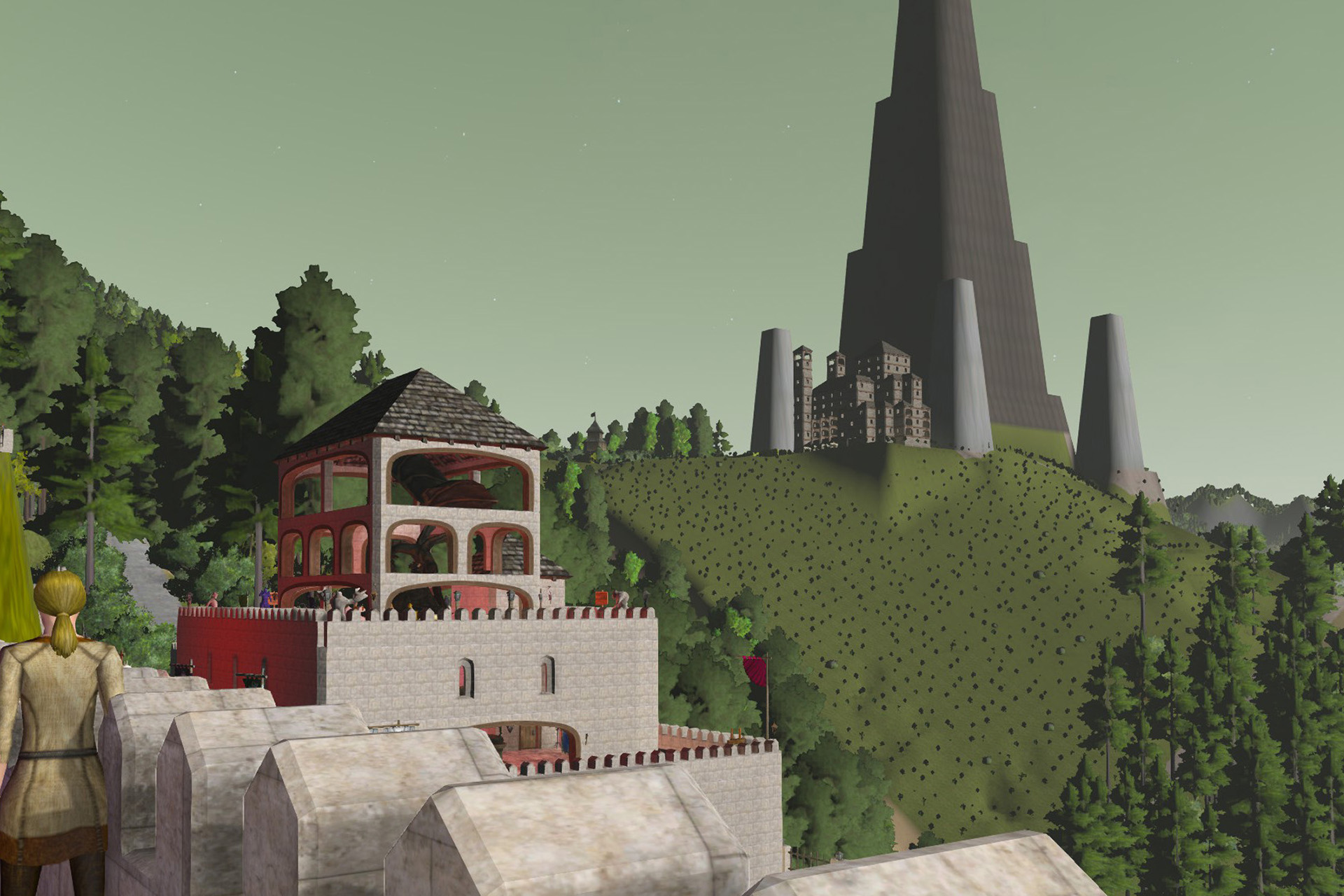}
    \caption{R1 on the roof of the Rockcliff Museum. In the distance, Fang Henge can be seen towering over the horizon. Just in front of it is the Rockcliff Cathedral.}
    \label{fig:wurm_screenshot_6}
\end{subfigure}
\hfill
\caption{Screenshots of go-along locations. All screenshots are from the perspective of R2, except Figure \ref{fig:wurm_screenshot_4}.}
\label{fig:screenshots}
\end{figure*}

\subsection{Go-alongs}
The “go-along” is a qualitative methodology in which a researcher literally ‘goes along’ with a participant, accompanying them on an everyday walk while observing and asking questions. Kusenbach describes the go-along as a hybrid between traditional participant observation and interviews, with the key difference being that the “ethnographers are able to observe their informants’ spatial practices in situ” \cite{kusenbach2003}. Furthermore, Kusenbach identifies that the go-along is particularly well-suited to studying how subjects engage with their environment, and the social architecture of specific communities \cite{kusenbach2003}, which makes it particularly suitable to exploring those dynamics in Wurm Online. 

To date there have been limited examples of go-alongs being conducted in digital space. Jørgensen defines the “media go-along” as a method in which the researcher and participant navigate social media together \cite{jorgensen2016}.  We follow Jørgensen’s approach of critically reexamining the affordances of the methodology in a digital context, as well as diverging from Kusenbach’s original approach in recognising the value of a more “contrived” go-along in which the researcher pre-determines some aspects of the interview \cite{jorgensen2016}. A more recent go-along study conducted in a VR environment also took this approach, having a checklist of points to direct the focus of the interview \cite{vindenes2021}. 

Though there has been limited academic study of \textit{Wurm}, and little application of the go-along in games, arguably there is an example of both in games media with a "ridealong" in \textit{Wurm} published in \textit{Rock Paper Shotgun} \cite{rps}. Furthermore, machinima that document players' personal histories with game spaces, such as Gina Hara's \textit{Your Place or Minecraft}\cite{hara}, arguably have affinities with the go-along.

\subsection{Ethical review and participant selection}
This study was subject to King's College London's Low Risk Review process\footnote{Ethical review reference number: LRS/DP-23/24-39970}. Participants were provided with an Information Sheet and Consent Form stating that the research team would publish anonymised transcripts of the go-alongs.  We later requested a modification to our project requesting participant consent to publish their in-game usernames, to have their role in the research be recognised and recorded should they wish for that. Two of the participants consented to this, while one did not. 

We posted invitations to participate in an interview on both the official \textit{Wurm} forums and the official Discord server on November 13th 2023. The posts clearly identified our interest in the Dragon Fang Pass on Independence, and a desire to hear from people who had used, constructed or maintained it. 


During the course of recruiting interviewees we were also made aware of the existence of the Rockcliff Museum, located a short distance from the Dragon Fang pass. Due to the archival nature of the project, we approached the curator of the museum and asked them if they would be willing to participate in an interview as well, to which they agreed.

\subsection{Interview structure}

The go-alongs were conducted within \textit{Wurm Online}, with voice chat enabled through Discord. Both the in-game activities and audio of our conversations were captured to aid subsequent transcription. The interview was structured in three parts, with pre-interview questions, the main go-along, and post-interview questions. The pre-interview questions pertained to what gaming setup participants used and whether they had a specific schedule for when they played, in acknowledgement of the wider “assemblage of play” \cite{taylor2009}. With some open-ended questions prepared ahead of time, the go-along constituted a semi-structured interview. The decision to have post-interview questions asking the participants to reflect on their experience of the go-alongs followed Moran’s practice of asking follow-up questions \cite{moran2022}. We also created maps of the go-along routes following the example of a study on memory mapping and go-alongs in a post-mining landscape \cite{granados2020}, which are included in this paper's appendix. 

The interview transcripts have been published on the Internet Archive \footnote{https://archive.org/details/wurm-online-go-along-transcripts}. Usernames of other players mentioned by participants were redacted, as this is identifiable information under GDPR. This presented a tension between our desire to accurately record the history of \textit{Wurm Online}, and the need to consider the ethical implications of publishing those histories. We also honoured requests by participants to redact sections of their interviews, and chose to redact the co-ordinates of an abandoned deed in order to protect it. The usernames of deceased players have not been redacted as UK GDPR only applies to the information of identifiable living people \cite{gdpr}, and we wanted to include their contributions to the game's community.

\subsection{Reflexive thematic analysis}

We conducted a reflexive thematic analysis on the go-along data \cite{braun2006}\cite{braun2019}. This specific type of thematic analysis was chosen as we wish to acknowledge our role in the process of knowledge production, following Braun and Clarke’s assertion that:
\begin{quoting}
Themes are creative and interpretive stories about the data, produced at the intersection of the researcher’s theoretical assumptions, their analytic resources and skill, and the data themselves. \cite{braun2019}
\end{quoting}
We take a constructivist epistemological approach, conceptualising that language is implicit in the construction of meaning, with an experiential orientation in that we focus on participant’s reporting of their own reality \cite{byrne}. We primarily aimed for an inductive analysis of the data, although we acknowledge that our approach was framed by preconceived research questions. Similarly, a combination of latent and semantic coding was used. One researcher coded the data, then discussed initial themes with the co-author, before iterating on them, following a collaborative and reflexive approach \cite{byrne}. The finalised thematic map can be seen in Figure \ref{fig:maps}, in the Appendix.

\section{Results}
\subsection{Summary}

We conducted go-alongs with three participants (abbreviated for consistency to P1, P2 and P3; the authors are referred to as R1 and R2). The paths that we took with each participant can be seen in Appendix A. Nirav (P1) is the curator of the Rockcliff Museum, Gumbo (P2) is associated with the creation of the Dragon Fang Pass and P3 is a long-term \textit{Wurm} player familiar with the Pass' history. The go-along with P1 lasted approximately two hours, one hour with P2, and an hour and a half with P3. P2 no longer has access to the game and did not want to use voice chat, so we compromised by conducting the interview entirely through text chat, using a YouTube video of Dragon Fang Pass \cite{dragonfangvideo} as a multimedia aid. Arguably, this does not constitute a digital go-along, however we believe that the interview is useful as a point of comparison with the other two.

\subsection{Themes}
\subsubsection{Distributed Identity}
Distributed Identity is a theme that encapsulates how participants understood themselves, and other players, as having identities that crossed not only different platforms but multiple game avatars as well. For example, P1 has five ‘toons’ in \textit{Wurm Online} (`toon' is slang for a character or account in an MMO), one of which exists on a separate PvP (Player vs Player) server to the one we were doing the go-alongs on:

\begin{quoting}
\noindent P1: I live in a cave by myself on Gold Coast, and I used to get killed, and at this point after seven years pretty much every kingdom knows me, and I can probably go anywhere I want\end{quoting}

\noindent P1’s identity as a museum curator in the game transcends individual avatar identities, allowing them freedom of movement even on servers where that would usually not be the norm. P3 chose to come to the go-along interview as the character that they had when they originally started playing on the Independence server in 2012:

\begin{quoting}\noindent R2: So this is - is this your usual character that you play on?

\noindent P3: No, this is my original character... the one that that was around when this [the Pass] was built.\end{quoting}

\noindent P3’s “original” toon was associated with a part of their identity tied to a particular time and place within \textit{Wurm}, a point with particular relevance to RQ2. P3 in particular also commented on how the server infrastructure of \textit{Wurm} has changed over time:

\begin{quoting}\noindent P3: Well, for one thing, Wurm was a lot more populated back then. When this tunnel was built this [Independence] was the only server.\end{quoting}

\noindent Arguably, the identity of the game \textit{Wurm Online} has become further fragmented over time through the segmentation of the game on different servers, leading to a different play experience, especially for those who remember the game before those changes were made.
Beyond \textit{Wurm} itself, participants commented on how the \textit{Wurm} forum is an important platform for reinforcing player reputation and enabling networking:

\begin{quoting}\noindent P3: Because he would be posting on the forums, and... that's a lot of name recognition right there\\
\noindent\rule{2cm}{0.4pt}\\
\noindent P2: Knew [redacted] from streaming, I think. Like at an Impalong event or some such thing. Thought popped in my head to ask him, via the game forum, and he was there next day. Awesome person.\end{quoting}

The \textit{Wurm} forum is a place where “name recognition” can be garnered through repeated posting. The sense of identity being tied to names is also linked to how items in Wurm will bear the signature of the person who crafted them, and it is these signatures that P1 particularly values in terms of their curation strategy:

\begin{quoting}\noindent P3: …what I'm looking for, I don't... really care if it's valuable or not, you know, the signatures are what matter, so I'll go through and I'll look at what signatures are there and [what matters is] whether I have those signatures and how big the items are and whether they're gonna fit in the museum at this point.\end{quoting}

\noindent The implication that signatures are particularly valuable from a preservation point of view relates to RQ3, and demonstrates how the affordances of distributed identity in \textit{Wurm} influence that.

\subsubsection{Presence and absence}

This theme pertains to our own experience of the go-along alongside our participants, how that shaped knowledge production, but also participants’ own ambivalent emotional relationship with \textit{Wurm} as a digital space. R1 accessed \textit{Wurm} through a Steam Deck, which presented difficulties due to \textit{Wurm}’s interface being designed for a desktop keyboard and mouse set-up. Several times R1 had difficulty accessing ladders and a cart during the go-along with P1, leading to a running joke:

\begin{quoting}\noindent R1: Yeah I chose the wrong ladder -laughing-\\
\noindent R2: Yeah you chose poorly!\\
\noindent R1: I did choose poorly -laughing-\\
\noindent P1: -laughing-\\
\noindent\rule{2cm}{0.4pt}\\
\noindent P1: You guys'll totally know how to do ladders at the end of this tower.\\
\noindent\rule{2cm}{0.4pt}\\
\noindent R1: Sorry it's a strange -inaudible- thing to like... I'm trying to use the touch screen...\\
\noindent\rule{2cm}{0.4pt}\\
\noindent P1 [joking]: I'm making a rule you can't disembark [from the cart].\end{quoting}

\noindent Having difficulty with simple actions was embarrassing, but also led us to reflect on lag as a concept that can be applied not only to technology, but also to experience -- we lagged behind our participants in terms of our own expertise with the game and its mechanics. This also became apparent when we encountered a hostile `cave bug' during the go-along with P3, leading to an awkward situation in which they prompted us to help them fight it:

\begin{quoting}\noindent P3: ... uh there's a cave bug. I can probably take it on, cos even though I'm not premium I do have gear that's pretty good. [P3 starts fighting the Cave Bug] You can help if you, if you want to...\end{quoting}

\noindent The affordances of the go-along, requiring us to traverse a route in-game with our participants and contend with our discrepancy in terms of lived experience in \textit{Wurm}, is very relevant to RQ1. In contrast, this kind of spontaneous experience was not possible during P2's interview. That being said, P2 felt that they didn’t even need a YouTube video of Dragon Fang Pass as a visual aid:

\begin{quoting}\noindent P2: time has flown by and didn't need the stream.. been through that tunnel more than enough times before\end{quoting}

\noindent This indicates that while participants believe they can rely on their own memories of a game, the texture of the interview responses will be different. This is demonstrated by P1 reflecting that:

\begin{quoting}\noindent P1: if we're having a phone call and I'm remembering stuff, and I'm answering your questions, I'm focused on your questions and we're not experiencing it, you know?\end{quoting}

\noindent Subthemes for the presence and absence themes are immateriality and emotion; these were identified in response to an apparent dissonance between an emotional attachment to the game and it's immaterial nature:

\begin{quoting}\noindent P2: all good... is only a game... pixels... memories always great\\
\noindent\rule{2cm}{0.4pt}\\
\noindent P1 [reflecting on an emotional response]: What am I \textit{thinking}, you know? I mean these are pixels, like, wh-, 'sacred ground', what am I even \textit{talking} about, I mean part of me was just like, am I getting too involved?\end{quoting}

\noindent P1 and P2 reflected on the nature of \textit{Wurm Online} as “pixels,” and the perception that the digital nature of the game would apparently make any kind of significant emotional response inappropriate. However, each participant arguably preserves their experience in their own way, whether through personal memory or in-game curation practises.

\subsubsection{Static vs living heritage}

We initially focused on Dragon Fang Pass as a case study for this work due to its designation as a Heritage Site by the community of \textit{Wurm Online}. Discussing the nature of heritage sites with our participants, we found that there were differing attitudes towards Heritage Sites with infrastructural affordances, and those that constituted a memorial or architectural monument:

\begin{quoting}\noindent P1: Um, so you have heritage sites that you know people made this really cool thing and it has meaning for the game, and you have all the canals and roads and boatways\end{quoting}

\noindent P3 explicitly encouraged us to look into community events rather than Heritage Sites like Dragon Fang Pass:

\begin{quoting}\noindent P3: [Dragon Fang Pass] it's essentially something that will remain here until the game shuts down and it's not gonna change much, but something like an impalong [a large community gathering], that is all player driven and you know interactive, and you can have like a hundred people connect into one.\end{quoting}

\noindent P2, who was involved in the construction of the Pass, felt that their involvement in a community event recreation of a TV programme was more significant than the tunnel:

\begin{quoting}\noindent P2: heh... Deal or No Deal is prolly more my "legacy"\end{quoting}

\noindent This theme encapsulates the tension between the desire to preserve content in the game in a way that renders it static, versus an interest in dynamic community events which cannot be captured through a traditional preservation strategy. The affordances of the Heritage Site designation in the game are actually inappropriate for P1's museum:

\begin{quoting}\noindent P1: we can't figure out how to make [the museum] a heritage site, and the reason why is because it's ever-changing, heritage sites usually lock 'em down and that's it, and then nothing can either be added or taken from it, and this has stuff that's on loan, and, it's gotta constantly be repaired and maintained, and... so I run it as a heritage site but it's not technically a heritage site.\end{quoting}

\noindent Furthermore, the official designation of Heritage Sites is only applicable to public-facing infrastructure or monuments. P3 maintains a settlement that they had with their friends as the only remaining inhabitant:

\begin{quoting}\noindent P3: Uh, well. You know, I can... I.. it was such a ... community with the friends and all, at the time, that I kind of like maintain it as a heritage site on my own. \end{quoting}

\noindent Attitudes towards the curation of content in \textit{Wurm} is context dependent; there may be a desire to engage with the living heritage of public community events with a coexistent desire to maintain a personal settlement as it once was. This indicates there is no singular answer to RQ3.

\subsubsection{Freedom and control}

Freedom and control is a complementary theme to static versus living heritage, however it is distinct in that it concerns the wider affordances of \textit{Wurm} and how that affects player experience. Participants expressed that the freedom to behave and interact with the game environment is what sets \textit{Wurm} apart:

\begin{quoting}\noindent P1: we have the freedom to be really amazing beautiful people, but we also have the freedom to be shitheads. And, you know, you can be who you... whoever you want to be in this game, but there's consequences and they're real\\
\noindent\rule{2cm}{0.4pt}\\
\noindent P3:  I... see Wurm as retaining the players it has now due to its uniqueness, you can't really find a game that has the same, you know, destructibility of environments, when everything is created by the players. \end{quoting}

\noindent This freedom to act altruistically or selfishly renders acts of generosity even more meaningful in the game because they are voluntary and not explicitly incentivised. P3 wanted to take us to The Howl, a starter town on the Independence server where new players had permission to stay without charge:

\begin{quoting}\noindent P3: The person that settled down here that I was helping out, that was a newer player, um that I had met at the Howl one time that I was just, you know, just passing through\end{quoting}

\noindent The designation of spawn points for new players affected the development of such friendships as well as permissions around infrastructure, as P2 chose for veins of valuable metal ores exposed by the creation of the Dragon Fang Pass to be publicly accessible:

\begin{quoting}\noindent P2: I left them open to mine because it's just helpful for new players to have access to that stuff. The spawn city was very very close, south side of Freedom Market.\end{quoting}

\noindent There is a culture of helping induct new players into \textit{Wurm}, and this is arguably influenced by the acknowledgement that the game is difficult due to the lack of control players have over the outcome of in-game mechanics:

\begin{quoting}\noindent P1: there's certain personalities that just cannot handle true random, there's some that don't like the step by step tediousness, the real life feel of this. \end{quoting}

\noindent The simultaneous freedom and lack of control that characterise \textit{Wurm} are important considerations for RQ1, while the significance of The Howl as a primer for discussion about helping new players is relevant for RQ2. In terms of curatorial practises, P1 describes an ostentatious candelabra made out of a precious metal they discovered:

\begin{quoting}\noindent P1:  And it was called "I am gay and dumb" which - I'm part of the LGBTQ community, and I wanted to leave that there because [laughing] I was like [laughing] this is fabulous, you know?\\
\noindent R1: Yeah, it is [laughing]\\
\noindent P1: But there are things that you're not able to put in a museum. Um. \end{quoting}

\noindent P1 has the freedom to decide how and what is preserved in the museum, and there are concerns around how objects will be interpreted without additional context -- this would be beyond their control.

\subsubsection{Tending}

Although somewhat unorthodox for a reflexive thematic analysis, we have a theme which could be defined as an overarching theme: tending. This theme links to all the others while still being distinct and internally consistent (see Figure \ref{fig:maps} in the Appendix for additional context). Under the themes of both tending and power and control are examples of participants looking after rare creatures in pens for a long period of real-world time. P3 is the last person left at their deed and must log into the game to feed five giant champion dogs at least every three days, and has been doing this for years:

\begin{quoting}\noindent P3:I have five dogs at least, five champion dogs at least, so that's five pumpkins a day, and the bucket holds twelve, so... like every three days I'd have to refill. \\
\noindent R1: Mmm, wow.\\
\noindent P3: Refill with pumpkins. That's not really a thing that we do as a community, I'm the only one left at my deed, so it's not like any of my friends are still there. Uh. \end{quoting}

\noindent P2 tells the story of a dragon that they looked after for five years before it escaped and was killed:
\begin{quoting}
\noindent P2: [Redacted] slowly drifted from the game... I kept up the dragon feedings and such... One day the in-escapeable pen.. Faultered\\
\noindent We knew she escaped and was loose for maybe a year... then she showed up and tore up a village, well they had a dragon slaying event. Was huge \end{quoting}

\noindent Here we have cases where players invested time and energy over a long period of time to maintain rare creatures, which could be conceived as a form of preservation relevant to RQ3. Another example of tending is in relation to friendships, and the presence and absence of those relationships. P3 showed us their boat at Freedom Docks, which had a Hungarian name:

\begin{quote}P3: I'm not, I don't... I don't even speak Hungarian, but... that was... kind of a community aspect that lasted. They were Hungarian, so you know, it was doing things like this to kind of show... uh... yeah to make it something they'd be familiar with and uh... be able to identify and... you know, be more comfortable with, I guess, would be a way of describing it.\end{quote}

\noindent Players tend to friendships through the affordances of the game and their ability to shape their environment. Tending activities also relate to the distributed identity of player signatures. As signatures deteriorate over time in \textit{Wurm}, in order for them to be legible a player will have to ‘imp’ or improve it, which requires repeated time and resource investment. The signatures are crucial for P1's curation of the museum, and thus:

\begin{quoting}\noindent P1: So the amount of time people have spent improving these items is uncountable, honestly. \end{quoting}

\noindent Tending as a repetitive activity was also reflected in the non-verbal actions of P3, who would stop to repair objects such as a fountain in Dragon Fang Pass during our go-along, without comment or drawing attention to it. These observations were one way in which the go-along methodology provided insight into how players habitually preserve \textit{Wurm} as they move through it. This persistent tending to \textit{Wurm} as a shared world is perhaps best expressed by P1:

\begin{quote}P1: Wurm is... going to go away. It's... that's life. This is not something that lasts forever, and... this is... this is a sacred place, the whole thing is, I'm not talking about the museum - the game. The game is a sacred space, it's a place where people... are... you know, we're \textit{here}, together. Making a world together.\end{quote}

The preservation of \textit{Wurm}, then, is a continuous group effort, and that itself is what lends it significance. 

\section{Discussion}

\subsection{Submerged play}

If the ethnography of MMOs which do not acknowledge the presence of ethnographers is termed “periscopic play,” then the go-along methodology could well be termed “submerged play.” To answer RQ1, we found that the go-along provided us with insights into the unique frictions of \textit{Wurm} as a game alongside our participants in ways that could be frustrating, amusing and  also deeply rewarding, allowing us further insight into their extensive lived experience in this digital space. Leaning into the affective qualities of the go-along \cite{stiegler2021}, whether that be getting frustrated with the Steam Deck or reflecting on a friendship that faded over time, provided us with a deeper context for the heritage of \textit{Wurm} beyond what was visible. There is a trend for digital ethnographers to self-validate by stating their own expertise with a game \cite{nicktaylor2008} but we seek to instead embrace the messy reality of our research and relinquish any kind of projected mastery over \textit{Wurm}, or indeed the direction of our go-alongs. Inviting our participants to guide the path we took led to discussions that we never could have predicted, which is apt for a game that is defined by freedom and the relinquishing of control. 

\subsection{Wurm as heritage site}

The affordances of \textit{Wurm} as a smaller scale MMO confirms T.L. Taylor’s speculation that:
\begin{quoting}Might there be ways – structural, economic, organizational, – in which smaller game worlds are at an advantage in exploring participatory practices, innovative forms of government, or even radical design challenges? \cite{taylor2006}\end{quoting}
The whole of \textit{Wurm}, as a landscape that has been shaped by players, developers and server infrastructure, is arguably a heritage site. With reference to RQ2, being in that landscape with participants contextualised their memories, but also the passage of time. For example, we followed P3 to the site of a friend’s former deed that had become overgrown in the last decade due to the designed decay of the game:

\begin{quoting}
\noindent P3: Okay, yeah it looks like this stretch over here of this woodland is where it's at, it's changed a great deal, it didn't used to be like it is now. But this plateau and where the trees are and all this is where his small deed was. \end{quoting}

\noindent Conducting our research of \textit{Wurm} in its present state, as an MMO that is arguably beyond its peak but still maintains a dedicated player base, has afforded us the opportunity to understand what it means for players to contend with a digital landscape that is simultaneously surprisingly persistent and painfully fragile.

\subsection{Preservation as process, not record}

With regards to RQ3, we found that players had different attitudes to the preservation of more personal heritage sites as opposed to the broader history of the game. P3 maintains their deed for no one but themself, while P1 is so renowned as a curator that this identity transcends any one avatar or server. P2 is content with memories, but is also barred from accessing the game itself. 

\textit{Wurm} demands not only time but repetitive tending from its players to be maintained. We heard of the huge time investment that goes into maintaining the memory of past players through improving signatures, and witnessed the habitual gesture of repairing monuments. Like the champion dogs that must be fed every three days, \textit{Wurm} is a tended garden that is constantly in tension between the desire to preserve it and the need for it to continue to grow for the sake of a living community. 

\begin{quoting}\noindent P1: I don't know if you saw Babylon 5 but I kinda see myself in that moment, Garibaldi shuts the lights down, he's the last guy there, you know? And... I feel like that's gonna be me. I'm gonna be the last person standing in this game when the lights go out. \end{quoting}

\noindent If indeed, ethnographic projects are never finished, only left \cite{walker}, then the process of preserving \textit{Wurm} will only finish when the last person leaves the game. It was never about saving a world, only finding a better way to live in it.

\begin{figure*}[t!]
    \includegraphics[width=0.8\textwidth]{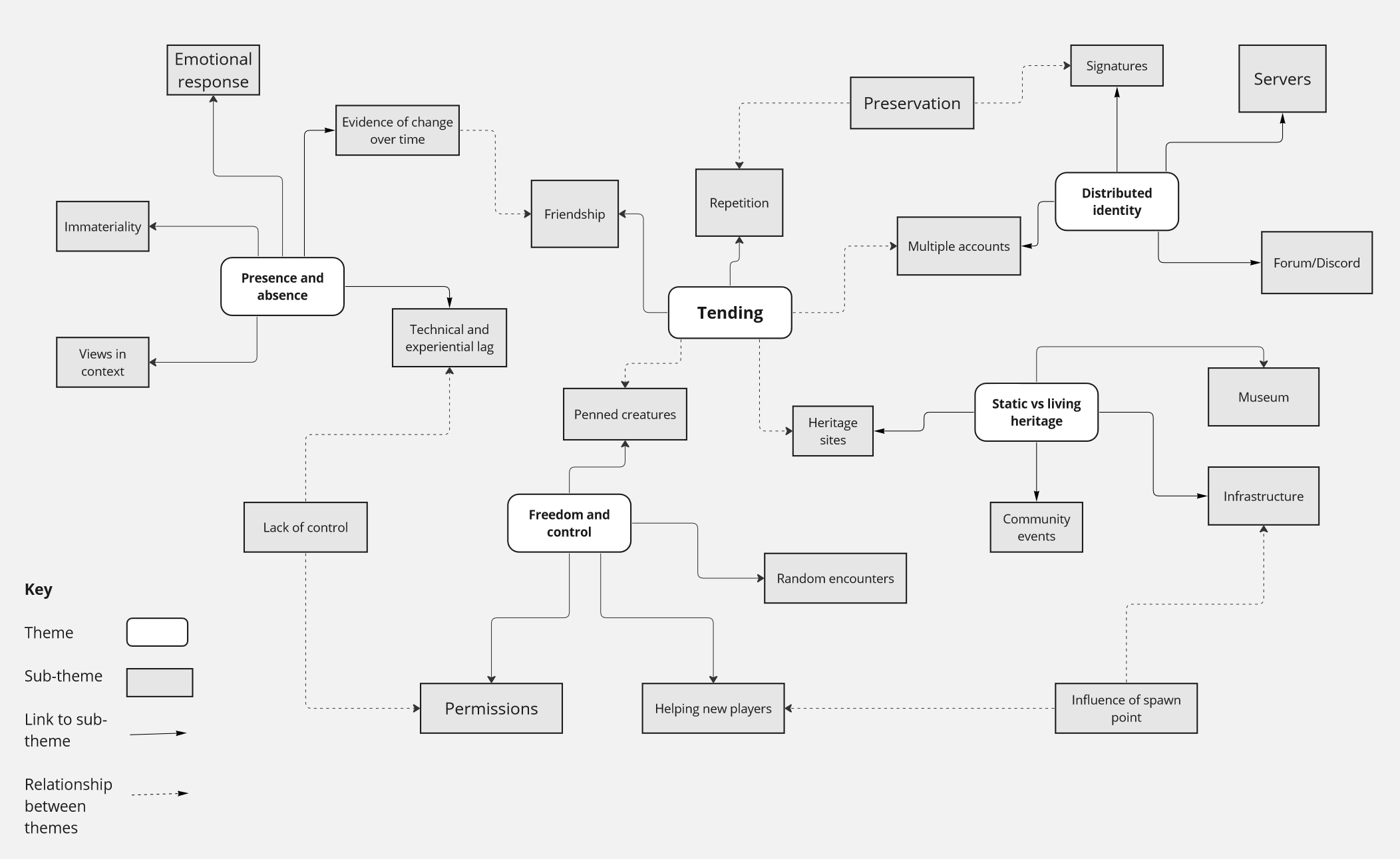}
    \caption{Finalised thematic map.}
    \label{fig:maps}
\end{figure*}

\section{Future Work}

\subsection{Wurm as living heritage}

While this study primarily focused on \textit{Wurm} player’s situated memories, the game’s community remains alive and active today, with a recent developer report even suggesting the player base is growing. As suggested by Participant 3, recurring events such as `impalongs' (where players `imp', or improve, one another's items and tools) would provide the opportunity to study active community events and document the living rather than just the static heritage of the game. Furthermore, participants referred to the differing cultures and attitudes between PvE (player versus environment) and PvP (player versus player) servers in \textit{Wurm}. One participant offered to negotiate permission for us to explore a PvP server safely, without the local player kingdoms attacking us (as new players are often mistaken for spies). Such permission may not extend to our interview participants, however, adding new complications and potentially interesting findings for the methodology. Studying and understanding both PvP and PvE communities in a game such as \textit{Wurm} is important to capture a more well-rounded perspective on the game's culture and history.

\subsection{The go-along as transcription}
In the process of transcribing the interviews, we observed that simply recording the verbal conversation was not sufficient to capture the go-along experience. For example, the tempo of conversation, interruptions and silences are also vital to understanding the dynamics of the discussion. As Stiegler puts it:

\begin{quote}Approaching “data” from a different perspective, the feelings that build up inside the actors in a research process—whether participant or researcher—coalesce into objects that necessitate examination rather than having these moments be interpreted as impediments to seeing or understanding the “real” data that was spoken into a tape recorder to be transcribed and analyzed later. \cite{stiegler2021}\end{quote}

Though we were of course not physically embodied within the game space for the go-along, non-verbal communication should also be considered in a digital context, whether that be through interaction with items or even the distance between different player avatars as they traverse a space. We believe that developing new approaches to visualisation and transcription of digital go-alongs would help record this information, leading to richer qualitative research, especially for future readers who may have a reduced familiarity with the contemporary game context.

\subsection{The go-along as record}
Our application of the go-along methodology in \textit{Wurm Online} was, in part, as a game preservation strategy. However, as mentioned above, more work can be done to adapt the methodology for a digital game context. Much has been written on the potential for recording gameplay footage for game preservation purposes \cite{prax2019more} \cite{nylund2015walkthrough}, and a video recording of go-alongs would allow for the capture of more nuance beyond that which a traditional transcript would provide. Furthermore, adopting a connective ethnography approach, as demonstrated by Pellicone and Ahn \cite{pellicone2018building}, would allow us to engage with the multiple communication platforms that \textit{Wurm} players use concurrently during play. This also ties in with conversations around the sustainability of game preservation that contends with both tangible and intangible cultural heritage \cite{garda2020cultural}. Newman and Simons \cite{newman2018game} have stressed the importance of future access to gameplay recordings that can provide a sense of what it was like to play a game in a specific historical context, which cannot be provided by merely maintaining access to the software itself. Though there would be attendant ethical and privacy considerations, video recordings of go-alongs could also include footage of the physical context in which the researchers (and potentially participants, if they consented) accessed the game. This has parallels with the work of the Popular Memory Archive \cite{stuckey2015remembering}, for example, which has collected photographs of players in the 1980s in order to document the domestic context of play in Australia and New Zealand at that time.

\subsection{Wurm as memory craft}
As mentioned above, \textit{Wurm} invites its players to tend to it with repetitive actions. Sullivan et al have observed that although many games include crafting mechanics, few capture the materiality, creativity and communal bonding of physical crafting communities \cite{sullivan2018games}. We would be interested in examining community events in \textit{Wurm}, such as impalongs, through the lens of craft, especially as a gendered practice. Evidence of craft practise has been widely studied in the archaeological record, and there is potential to study the affordances as craft as social memory in \textit{Wurm} as a form of contemporary archaeology. As Fulcher puts it:

\begin{quote}Material metaphors can be physically experienced, and engagement with a material solidifies this conceptualisation, making the metaphor real. For example, walking through a door is an analogy for transition in many cultures; the experience not only expresses the concept but also help us to understand it \cite{fulcher2019practising}\end{quote}

Thus, in order for us to understand \textit{Wurm} as a community crafted space, it is imperative our documentation go beyond merely recording the landscape, we must also engage with it as a space that has personal and collective mnemonic affordances.

\section{Conclusion}

In this exploratory study, we found that applying the go-along to the digital space of the MMO \textit{Wurm Online} provided us with incredibly rich qualitative data, particularly with regards to how specific locations elicited memories and reflection. We interviewed three participants with diverse experiences in the game, and then conducted a reflexive thematic analysis. We identified five themes, with the overarching theme of tending arguably linking the other four. The go-along invited us to reflect on our own role and limitations as researchers, and to challenge the idea that to preserve something we have to freeze it in time. As \textit{Wurm} constantly invites its players to reaffirm themselves through repetitive actions, no static record could ever encapsulate its \textit{genius loci}.\\

\begin{acks}
We would like to thank Dr Sam Stiegler, who introduced the first author to the go-along method at the inaugural Hunt-Simes Institute in Sexuality Studies at the Sydney Social Sciences and Humanities Research Centre in 2023. 

We also wish to thank Nirav, Gumbo and Participant 3 for taking part in the project and being so generous with their time. We would also like to acknowledge the Wurm Online community, their fora and archives which were invaluable for this work.

The second author would also like to thank the community of players with which they first played Wurm Online many years ago, and who continue to be wonderful friends today. Your collective willingness to spend two weeks of your life trying and failing to make clay bowls with them is at least part of the reason this paper was possible at all.

Finally, our sincere thanks to the reviewers for their insightful suggestions and kind feedback. 

This work was supported by the EPSRC Centre for Doctoral Training in Intelligent Games  \&  Games Intelligence (IGGI) [EP/L015\-846/1] and the Royal Academy of Engineering.
\end{acks}


\newpage

\bibliographystyle{ACM-Reference-Format}
\bibliography{sample-base,goalongs}


\newpage

\appendix
\section{Appendix} 
\begin{figure*}[t]
    \includegraphics[width=\textwidth]{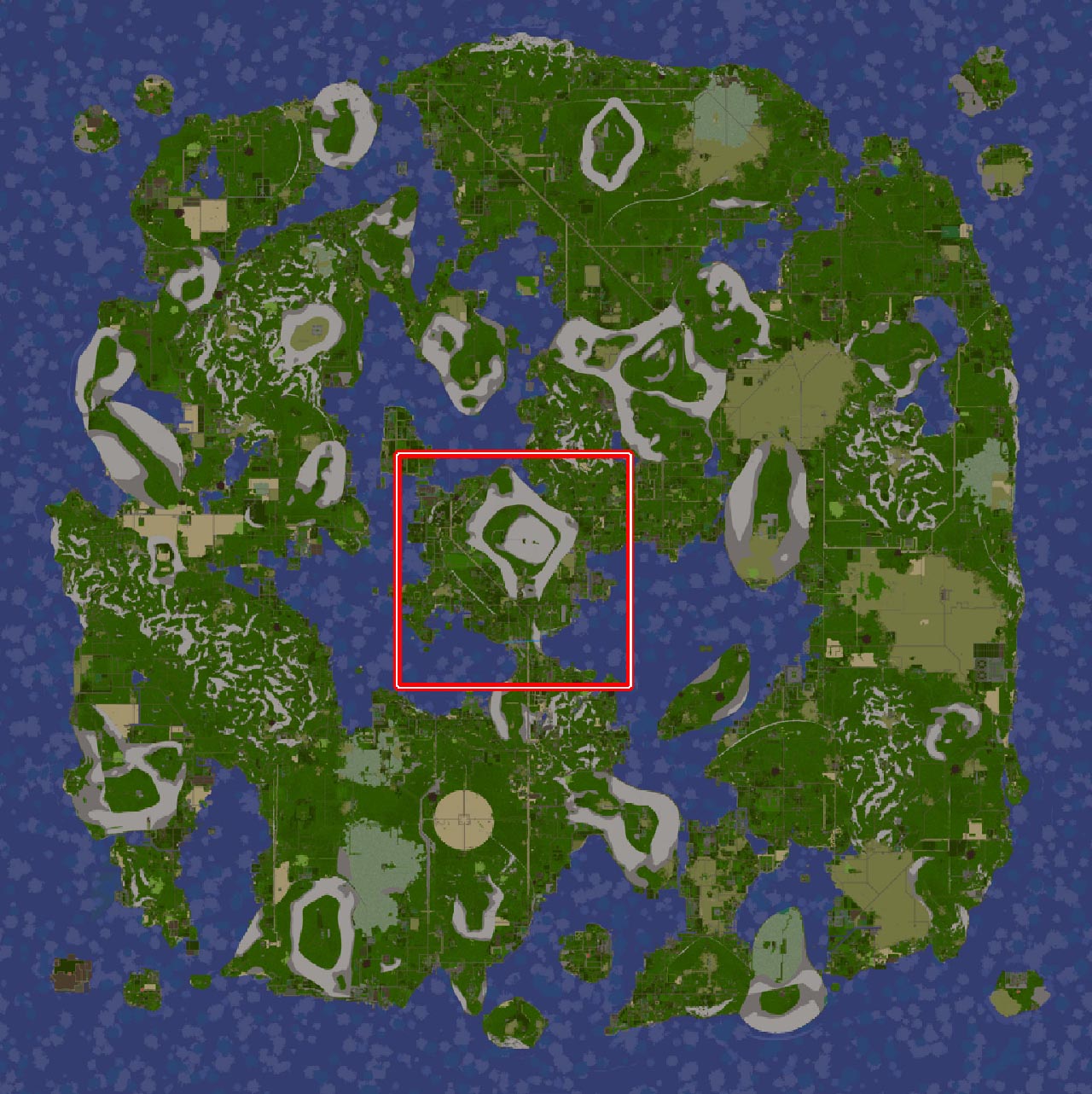}
    \caption{A player-created map of Independence\cite{wurmmap}, in full. The highlighted square is the region Figures \ref{fig:nirav2}, \ref{fig:gumbo2} and \ref{fig:p3} are based on. Travelling from the north coast of the map to the south coast, on foot, would take approximately 4.5 hours assuming consistent road coverage (based on a traversal of Cadence, which is a Wurm map of a similar size \cite{howbigmap}).}
    \label{fig:colormapfull}
\end{figure*}

In this appendix we provide maps detailing the area around Dragon Fang Mountain that we traversed during the go-alongs. Figure \ref{fig:colormapfull} shows a full player-made map of the Independence server\cite{wurmmap}, with the area we are focusing on marked. We also include three maps made by the authors, based on the community-created map, to mark each go-along path, as well as notes on key points visited throughout. The go-alongs for Participant 1, 2 and 3 are found in Figures \ref{fig:nirav2}, \ref{fig:gumbo2} and \ref{fig:p3} respectively. 

\subsection{Points of Interest}
\begin{figure*}[t]
    \includegraphics[width=\textwidth]{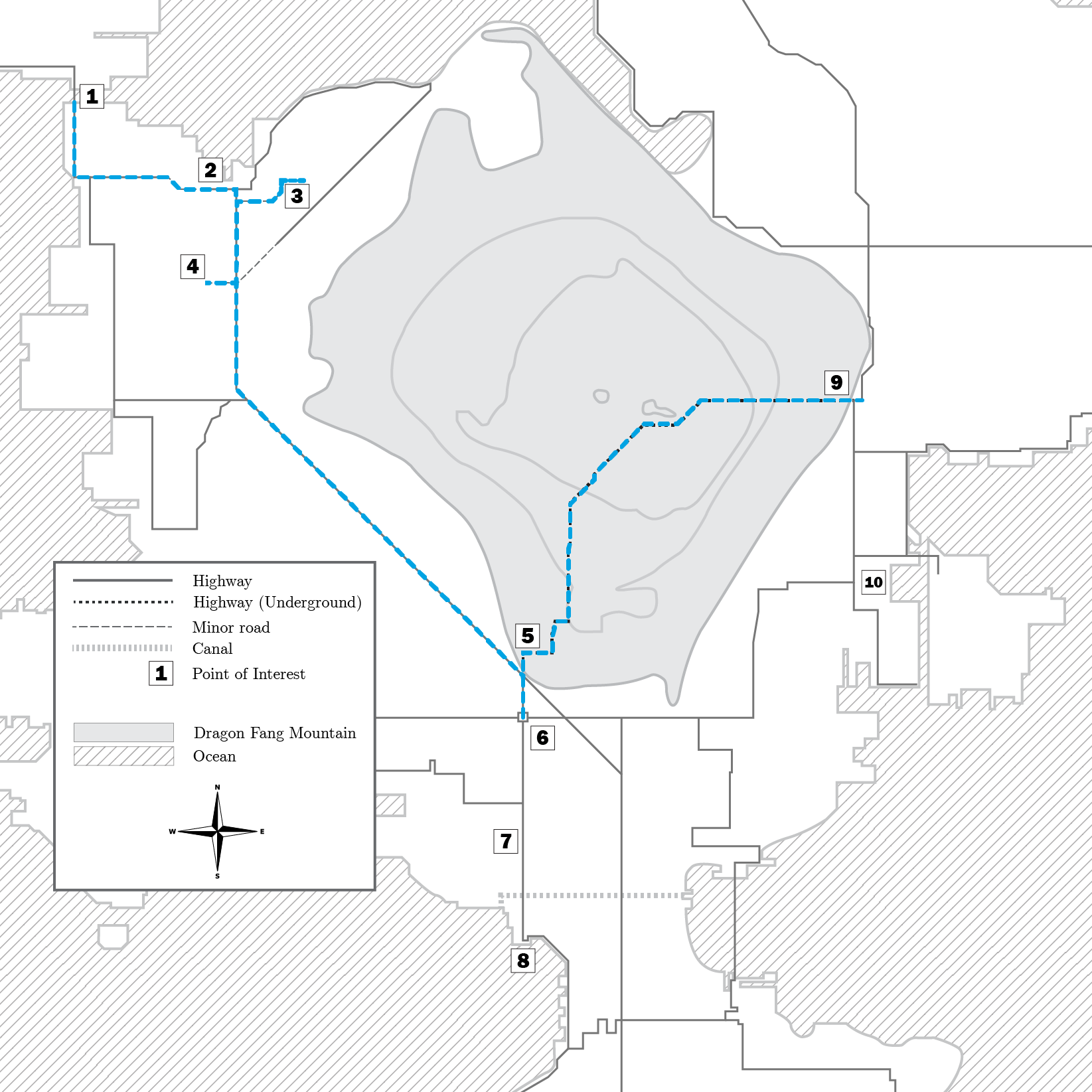}
    \caption{A map depicting the route taken on the go-along with Participant 1 (Nirav, curator of the Rockcliff Museum). The go-along began at the Rockcliff Museum (3) before beginning to travel via horse and cart to Fang Henge and the Rockcliff Cathedral (4). We then travelled to Lyric Beach (2) for supplies and visited the Black Dog Canal (1), continuing on to Freedom Market (6) before moving through the south entrance to the Pass (5) to the eastern entrance (9). We then retraced our steps back to Lyric Beach (2) to conclude.}
    \label{fig:nirav2}
\end{figure*}

\begin{figure*}[t]
    \includegraphics[width=\textwidth]{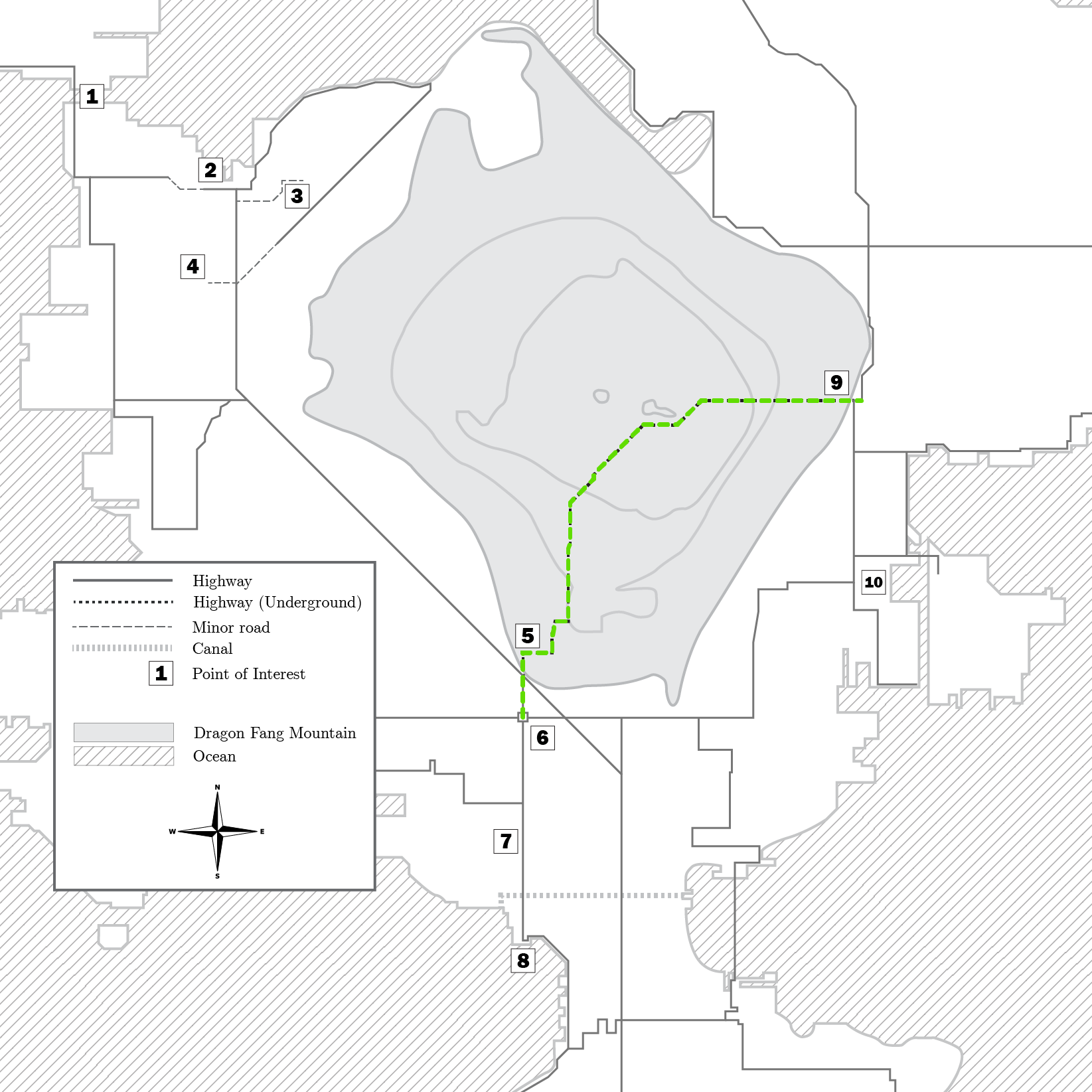}
    \caption{A map depicting the route observed via video during the go-along with Participant 2 (Gumbo, who is associated with the creation of the Dragon Fang Pass). The video can be accessed online \cite{dragonfangvideo}.}
    \label{fig:gumbo2}
\end{figure*}

\begin{figure*}[t]
    \includegraphics[width=\textwidth]{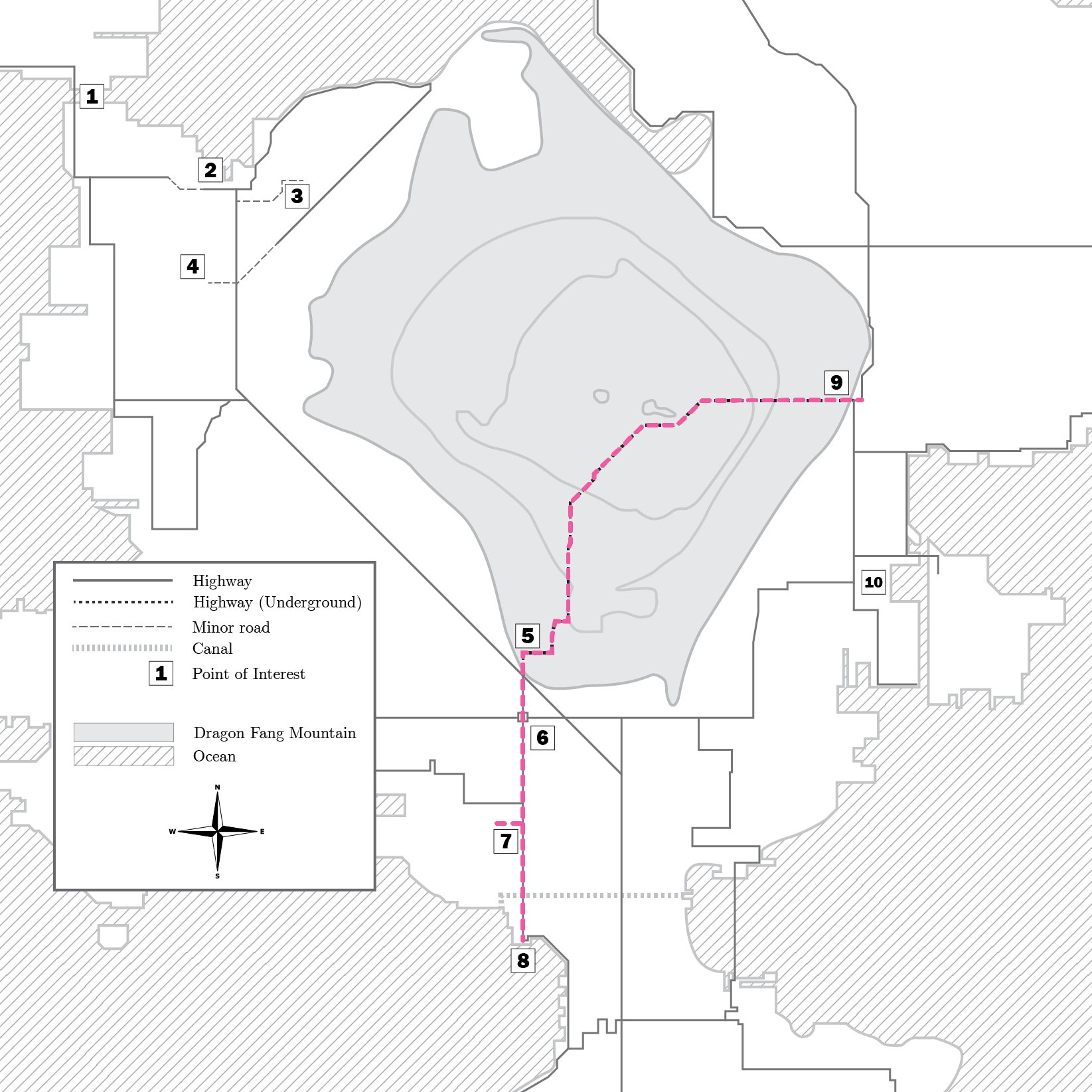}
    \caption{A map depicting the route taken on the go-along with Participant 3. The go-along began at Freedom Market (6), and led us down to Freedom Docks (8) via The Howl (7). We then returned to Freedom Market, passed through the south entrance to the Pass (5) and concluded our walk at the east entrance (9).}
    \label{fig:p3}
\end{figure*}

Several points of interest are marked on the maps in Figures \ref{fig:nirav2}, \ref{fig:gumbo2} and \ref{fig:p3}. The numbers on the labels relate to the points below. 


\begin{enumerate}
    \item Black Dog Canal. A heritage site. Canals are dug by players, and even a small canal such as this represents a major engineering feat. We believe that prior to the creation of the canal, Black Dog Island to the north was actually a peninsula attached to the same landmass as the Dragon Fang Mountain. The heritage site preserves the canal as an important thoroughfare for ships. 
    \item Lyric Beach. A town built and maintained by the curator of the Rockcliff Museum, built on the remains of an earlier deed owned by the creators of Fang Henge and the Rockcliff Cathedral (see below). Participant 1 took us to Lyric Beach to provide us with supplies and afterwards brought us back to Lyric Beach so that our characters could sleep at the inn overnight prior to conducting our later go-alongs.
    \item Rockcliff Museum. The site of the museum where our go-along with Participant 1 began. Figure \ref{fig:wurm_screenshot_2} shows a memorial inside the Wurm museum grounds.
    \item Fang Henge and Rockcliff Cathedral. Fang Henge is a player-made monument, now standing next to Rockcliff Cathedral, a heritage site. Both were built by famous members of the community, and their significance is such that in-game tapestry items can be crafted depicting their construction. Fang Henge dates back to 2010, while the Cathedral was constructed in 2013. The main architect, Tich, has since passed away.
    \item Dragon Fang Pass, South. This is the southern entrance to the pass, near Freedom Market.
    \item Freedom Market. Built by players, Freedom Market is a collection of market stalls which were previously staffed with NPC merchants selling player-made goods. Dozens of stalls still stand here, mostly abandoned. The Market's significance as a focal point for players is one of the reasons the Dragon Fang Pass was originally built, and why one of its entrances is so near the market.
    \item The Howl. When new players join a server in Wurm, they appear at one of the server's starter towns. The Howl is the only starter town currently active on Independence. It was not the original starter town for the server, but it became active in the early 2010s. This, combined with Freedom Market's draw as a trading hub, made this part of the server the focal point of Independence.
    \item Freedom Docks. A long highway connects the Dragon Fang Pass, Freedom Market, The Howl and the docks at the south end. Freedom Docks is the nearest accessible shore to Freedom Market, and therefore a convenient place for players travelling by boat to anchor their ships. A small harbour has been built here, which Participant 3 took us to see.
    \item Dragon Fang Pass, East. This is the eastern entrance to the pass. Several highways pass by this area, providing access further north and east across the continent. Figure \ref{fig:wurm_screenshot_1} shows one of the authors, R1, standing by the heritage site sign marking the entrance to the pass.
    \item A collection of player-owned deeds. One or more of these deeds were previously held by Gumbo, the player who led the construction of the Dragon Fang Pass. All participants commented on the view down the hill from the eastern pass entrance to this area, although no go-along visited it.
\end{enumerate}

\end{document}